\documentclass
[aps,prb,twocolumn,showpacs,amsmath,amssymb,superscriptaddress,floatfix]{revtex4}%
\usepackage{amsmath}
\usepackage{txfonts}
\usepackage{color}
\usepackage{graphicx}
\usepackage{dcolumn}
\usepackage{bm}
\usepackage{longtable}
\usepackage{amsfonts}
\usepackage{amssymb}%
\providecommand{\U}[1]{\protect\rule{.1in}{.1in}}

\begin{document}
\title{ Spin pumping into two-dimensional electron systems }
\date{\today}
\author{Takuya Inoue}
\affiliation{Institute for Materials Research, Tohoku University, Sendai 980-8577, Japan}
\author{Gerrit E. W. Bauer}
\affiliation{Institute for Materials Research, Tohoku University, Sendai 980-8577, Japan}
\affiliation{Kavli Institute of NanoScience, TU Delft Lorentzweg 1, 2628 CJ Delft, The Netherlands}
\affiliation{WPI-AIMR, Tohoku University, Sendai 980-8577, Japan}
\author{Kentaro Nomura}
\affiliation{Institute for Materials Research, Tohoku University, Sendai 980-8577, Japan}

\begin{abstract}
We study the spin current injected by spin pumping into single layer graphene
and the two-dimensional electron gas (2DEG) with ferromagnetic contacts using
scattering theory. The spin currents pumped into graphene are very distinct
from that into the 2DEG since importantly affected by Klein tunneling.

\end{abstract}
\maketitle

\section{Introduction}

Spintronics takes advantage of both the charge and spin degrees of freedom of
the electron to generate novel device functionalities for information and
communication technologies. Key concepts in spintronics are the spin current,
i.e. net flows of spin angular momentum, and the spin accumulation, i.e. a
non-equilibrium imbalance between the densities of the two spin species. Both
can be generated in non-magnetic conductors by several methods, such as
electrical \cite{ohno1999,jedema2001}, optical \cite{taniyama2011}, and
thermal spin injection \cite{jeon2013} and the spin-orbit interaction
\cite{maekawa2013}. It is also possible to inject spins dynamically into
various non-magnetic conductors by \textquotedblleft pumping\textquotedblright%
\ via a ferromagnetic contact with a time-dependent magnetization. The effect
can be understood in terms of adiabatic quantum pumping, i.e. the generation
of a current not by an applied voltage but a\ time-periodic modulation of the
scattering matrix by external parameters. Spin pumping does not suffer from
the conductance mismatch problem that plagues electrical spin injection
\cite{Brataas2002,Wu2014} and is at the root of many physical phenomena, e.g.
the spin Seebeck effect \cite{Uchida2010}. Here we report a theoretical study
on spin pumping into two-dimensional electronic systems, such as the
two-dimensional electron gas (2DEG) in layered semiconductors and graphene.

A 2DEG system may be formed at the interface between two different
semiconductors, such as modulation-doped GaAs$|$AlGaAs heterojunctions or
quantum wells that at low temperatures support electron mobilities of
$>10^{7}\,$\textrm{cm}$^{2}$\textrm{V}$^{-1}$\textrm{s}$^{-1}$ corresponding
to transport mean-free paths of $\simeq100\,\mathrm{\mu}$m \cite{Umansky}.
While spin injection into thin semiconductor layers has been reported
\cite{Lou2007},\ we are not aware of spin injection experiments into
high-mobility 2DEGs, presumably due to the conductance mismatch problem.

Monatomic layers such as graphene and transition metal dichalcogenides also
have two-dimensional electronic structures
\cite{katsnelson2006,ohshima2014,shao2014,tang2013} and are interesting candidates for
spintronic applications \cite{GS,Mendes,Drogeler}. For graphene high electronic mobilities of
$3.5\times10^{5}\,$\textrm{cm}$^{2}$\textrm{V}$^{-1}$\textrm{s}$^{-1}$ have
been reported \cite{Banszerus2015}. Electric spin injection has been achieved
with long spin flip diffusion lengths of$\ 13/24\,\mathrm{\mu m}$ at room
temperature/4 K, respectively \cite{Ingla-Aynes2015}. The low-energy
excitations of graphene close to the charge-neutral Fermi energy are well
described by the massless-Dirac equation \cite{neto2009} and to transport
properties very different from those of 2DEGs.

Recently, experiments on ferromagnetic resonance (FMR) of permalloy (Py) on
graphene report enhancement of the intrinsic resonance line widths and
non-local voltages \cite{tang2013, singh2013, patra2012}. These experiments
are interpreted as evidence for spin currents pumped into graphene
\cite{mizukami2002,tserkovnyak2005,tserkovnyak2002, tserkovnyak2002b}. Spin pumping can be
formulated \cite{tserkovnyak2002,tserkovnyak2002b,tserkovnyak2005} as a
spin-dependent generalization of the B\"{u}ttiker-Brouwer adiabatic quantum
pumping formula based on the scattering theory of transport \cite{brouwer1998}%
. Rahimi and Moghaddam \cite{rahimi2014} computed spin pumping into graphene
by a magnetic insulator, which has the advantage that parallel conductance
channels that may exists for magnetic metals are completely suppressed.
Recently, transport experiments of graphene on a magnetic insulator Yttrium
Iron Garnet (YIG) substrate have been reported \cite{GYIG,Wees2015} that show
an induced proximity exchange splitting of $\sim0.1\,$T in the graphene
electronic structure \cite{Wees2015}. Much larger exchange splittings have
been predicted for graphene on EuO \cite{EuO} and observed for graphene%
$\vert$%
EuS \cite{EuS}. We note that spin pumping into 2D systems by an electrically
insulating magnetic gate may efficiently emulate the spin pumping and maser
action predicted to occur by inhomogeneous Zeeman fields
\cite{Watts06a,Watts06b}.

\begin{figure}[th]
\begin{center}
\includegraphics[width=0.475\textwidth]{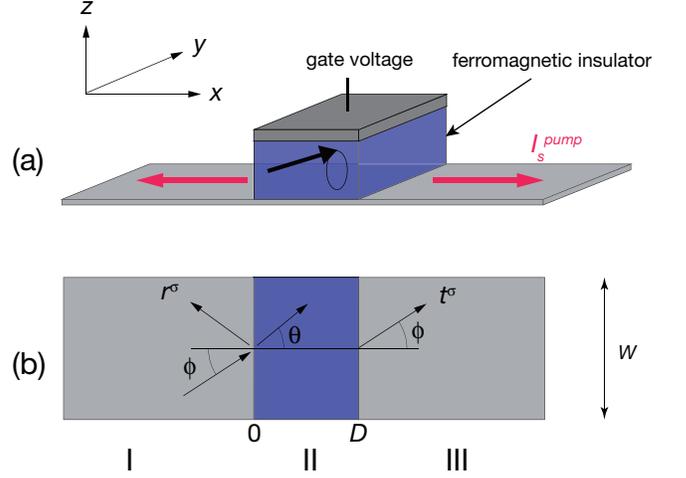}
\end{center}
\caption{(a) Schematic of the spin-pumping device: a metallic gate on a
ferromagnetic insulator film on top of a 2D electronic system, either a 2DEG
or graphene. We calculate reflection and transmission coefficients for
electrons impinging on the gated region. (b) $\phi$ is the angle of incidence
and $\theta$ the refraction angle. Red arrows represent the spin currents
induced by spin pumping.}%
\label{system}%
\end{figure}

Here we consider a ferromagnetic insulator (FI) on top of a two-dimensional
electron system (2DEG or graphene) connected to electron reservoirs as in Fig
\ref{system}, where the latter are kept at the same chemical potential. An
additional metallic gate on top of the FI tunes the electron density and Fermi
energy of the electrons relative to that in the reservoirs. The exchange
interaction of the conduction electrons with the ferromagnet induces a
proximity exchange potential that weakly magnetizes the electron gas
\cite{mcguire2004,wang2015,Wees2015}. When the FI magnetization moves
sufficiently slowly, \textit{e.g}. under FMR, the induced magnetization
follows adiabatically. The scattering matrix connecting the reservoirs changes
parametrically in time, thereby pumping a spin current into the reservoirs.
The spin coherence length $\lambda=\pi/\left\vert k_{F}^{\uparrow}%
-k_{F}^{\downarrow}\right\vert $, where $k_{F}^{\sigma}$ is the Fermi wave
number of the conduction electrons under the ferromagnet with $\sigma
=\uparrow$ or $\downarrow$, is now much larger than that of metallic
ferromagnets, in which $\lambda$ is of the order of a few \aa ngstr\"{o}ms.
Spin-dependent DC transport in such proximity-magnetized graphene has been
studied theoretically \cite{brataas2008,yokoyama2008}. Here we consider the
spin pumping of a weakly magnetized ballistic electron gas (2DEG and graphene)
with slightly different Fermi circles for up and down spins and a $\lambda$
that should be larger than the length $D$ of the scattering region.

In the absence of spin-orbit interaction, the scattering matrix for a
mono-domain ferromagnetic element sandwiched by two normal metals may be
decomposed as%
\begin{equation}
S=S^{\uparrow}\hat{u}^{\uparrow}+S^{\downarrow}\hat{u}^{\downarrow},
\end{equation}
where $S^{\sigma}$ is the scattering matrix for spins up (down) along
$\mathbf{m}$, the unit vector of magnetization of the ferromagnet. The
spin-projected scattering matrix\
\begin{equation}
S^{\sigma}=\left(
\begin{array}
[c]{cc}%
r^{\sigma} & t^{\sigma}\\
t^{\sigma} & r^{\sigma}%
\end{array}
\right)  ,
\end{equation}
where $t^{\sigma}/r^{\sigma}$ are transmission/reflection coefficient matrices
for spin $\sigma$. $\hat{u}^{\sigma}$ is the projection operator
\begin{equation}
\hat{u}^{\sigma}=\frac{1}{2}(1\pm\mathbf{\hat{s}}\cdot\mathbf{m}),
\end{equation}
where $\hat{\mathbf{s}}=\sum_{l=x,y,x}\hat{s}_{l}\mathbf{e}_{l}$ and $\hat
{s}_{l}$ are the Pauli matrices. The spin current pumped into adjacent normal
metals then reads
\begin{equation}
\mathbf{I}_{s,R}^{pump}=\frac{\hbar}{4\pi}(g_{r}\mathbf{m}\times
\frac{d\mathbf{m}}{dt}-g_{i}\frac{d\mathbf{m}}{dt}),
\end{equation}
where $g=\sum_{nn^{\prime}}(\delta_{nn^{\prime}}-r_{nn^{\prime}}^{\uparrow
}(r_{nn^{\prime}}^{\downarrow})^{\ast})-t_{nn^{\prime}}^{\uparrow
}(t_{nn^{\prime}}^{\downarrow})^{\ast}$ is the complex spin-mixing
conductance\ with $\operatorname{Re}g=g_{r}$, $\operatorname{Im}g=g_{i}$. We
focus here on wide two-dimensional systems with widths $W$ with continuous
transport channel index $n\rightarrow k_{y}=k_{F}\sin\phi$, where $k_{F}$ is
the Fermi wave number in the leads and $\phi$ the angle of incidence. In the
following we assume a ballistic scattering region; the wave numbers are
conserved and all matrices diagonal. Then
\begin{equation}
g=\frac{2k_{F}W}{\pi}\int_{0}^{\pi}d\phi(1-r^{\uparrow}(\phi)(r^{\downarrow
}(\phi))^{\ast}-t^{\uparrow}(\phi)(t^{\downarrow}(\phi))^{\ast}). \label{SPC}%
\end{equation}

The proximity exchange potential of the (single-domain) ferromagnet polarizes
the conduction electrons. Under ferromagnetic resonance conditions the
magnetization precesses around the $z$ axis $\mathbf{m}(t)=(\sqrt{1-m^{2}}%
\cos\omega t,\sqrt{1-m^{2}}\sin\omega t,m)$ where $m$ is the cosine of the
precession cone angle. Then the instantaneous spin current pumped into the
adjacent leads reads
\begin{equation}
\mathbf{I}_{s,R}^{pump}(t)=\frac{\hbar\omega}{4\pi}\left(
\begin{array}
[c]{c}%
m\sqrt{1-m^{2}}g_{r}\cos\omega t+\sqrt{1-m^{2}}g_{i}\sin\omega t\\
m\sqrt{1-m^{2}}g_{r}\sin\omega t-\sqrt{1-m^{2}}g_{i}\cos\omega t\\
(1-m^{2})g_{r}%
\end{array}
\right)  ,
\end{equation}
while its time average becomes
\begin{equation}
\mathbf{J}_{s}=\frac{\omega}{2\pi}\int_{0}^{\frac{2\pi}{\omega}}%
dt\ \mathbf{I}_{s}^{pump}(t)=\frac{\hbar\omega}{4\pi}(1-m^{2})g_{r}%
\mathbf{e}_{z}.
\end{equation}
When this spin current is dissipated in the conductor or reservoirs, the loss
of angular momentum and energy increases the viscous damping of the
magnetization dynamics that is observable as an enhanced broadening of
ferromagnetic resonances \cite{tserkovnyak2002, mizukami2002,tserkovnyak2005}.
The spin-current may be converted into a charge signal by metallic contacts
that either have a large spin Hall angle \cite{saitoh2006} or are
ferromagnetic \cite{Brataas2002}. In the following we focus on the principle
of spin current generation, but leave the modeling of the spin current
detection for future study.

\section{Model \& results}

We consider Hamiltonians of the form
\begin{equation}
H^{2D}=H_{kin}^{2D}+U(\mathbf{r},t),
\end{equation}
where $H_{kin}^{2D}$ is the kinetic energy of the electrons that experience a
spin-dependent potential below the FI-gated area:
\begin{equation}
U(\mathbf{r},t)=V_{c}-J\mathbf{m}(t)\cdot\mathbf{s.}%
\end{equation}
$V_{c}$ is the electric potential controlled by the metal gate and
$-J\mathbf{m}(t)\cdot\mathbf{s}$ is the exchange potential that parametrically
depends on the FI magnetization direction $\mathbf{m}$ and electron spin
$\mathbf{s}$. In the absence of more detailed information $J$ is taken to be
not depend on $V_{c}.$ At equilibrium, the magnetization direction
$\mathbf{m}_{0}\Vert\hat{z}$ is constant and the potential for up (down) spin
electrons along the spin quantization $\hat{z}$-axis reduces to
\begin{equation}
V_{\sigma}\equiv V_{c}-\sigma J
\end{equation}
with $\sigma=\pm1$ for spin up and down. The spin current per unit width
$j_{s}=J_{s}/W$ is a function of several parameters:
\begin{equation}
j_{s}=j_{s}\left(  E_{F},V_{c},J,D\right)  .
\end{equation}
Outside the gated region the potential vanishes. Its shape in the intermediate
region depends on the device design. Here we consider two limits. When the
potential varies slowly on the scale of the electron wave length, an adiabatic
approximation is appropriate \cite{reijnders2013}. In this model the
potentials changes slowly on the scale of the electron wave length from
$V_{\sigma}^{(\mathrm{slow})}(\mathbf{r})=0$ for $x\ll0$ to $V_{\sigma
}^{(\mathrm{slow})}(\mathbf{r})=V_{\sigma}$ for $0\leq x\leq D$ and then back
to $V_{\sigma}^{(\mathrm{slow})}(\mathbf{r})=0$ for $x\gg0$.
%
In the other limit the potential at the boundaries of the scattering region
changes abruptly (on the scale of the electron wave length):
\begin{equation}
V_{\sigma}^{(\mathrm{abrupt})}\left(  \mathbf{r}\right)  =\left\{
\begin{array}
[c]{ll}%
0, & \quad x<0\\
V_{\sigma}, & \quad0\leq x\leq D\\
0, & \quad D<x.
\end{array}
\right.  \label{Vapr}%
\end{equation}
The scattering at the step potential can be treated by quantum mechanical wave
function matching. The reality is likely to be an intermediate between the two
extremes and can be understood qualitatively by interpolation.

\subsection{2DEG}

First, we discuss a wide strip of a 2DEG. In the Hamiltonian
\begin{equation}
H_{kin}^{\mathrm{2DEG}}=-\frac{\hbar^{2}\nabla^{2}}{2m^{\ast}},
\end{equation}
$m^{\ast}$ is the effective mass. Assuming sufficiently wide strips we may use
the periodic boundary condition in the lateral $y$-direction. We consider
electrons that in the reservoirs are at the Fermi energy $E_{F}=\hbar^{2}%
k_{F}^{2}/\left(  2m^{\ast}\right)  $ with $k_{F}=\sqrt{k_{x}^{2}+k_{y}^{2}}$.

We first discuss the abrupt potential limit in which the Hamiltonian can be
written as
\begin{equation}
H^{\mathrm{2DEG}}=-\frac{\hbar^{2}\nabla^{2}}{2m^{\ast}}+V_{\sigma
}^{(\mathrm{abrupt})}\left(  \mathbf{r}\right)  .
\end{equation}
The electrons in region I ($x<0$) are a linear combination of incoming and
reflected waves (see Figure \ref{system})
\begin{equation}
\psi_{I,\sigma,\mathbf{k}}\left(  \mathbf{r}\right)  =e^{ik_{x}x+ik_{y}%
y}+r^{\sigma}e^{-ik_{x}x+ik_{y}y}.
\end{equation}
In region II ($0<x<D$)
\begin{equation}
\psi_{II,\sigma,\mathbf{k}}\left(  \mathbf{r}\right)  =a^{\sigma
}e^{iq_{x,\mathbf{{k}}}^{\sigma}x+iq_{y}^{\sigma}y}+b^{\sigma}%
e^{-iq_{x,\mathbf{{k}}}^{\sigma}x+iq_{y}^{\sigma}y},
\end{equation}
and in region III ($x>D$)
\begin{equation}
\psi_{III,\sigma,\mathbf{k}}\left(  \mathbf{r}\right)  =t^{\sigma}%
e^{ik_{x}x+ik_{y}y},
\end{equation}
where the indices $i=\left\{  I,II,III\right\}  ,\,\sigma=\pm,$ and
$\mathbf{k}$ denote electrons with spin $\sigma$ in region $i$ with wave
vector $\mathbf{k}=\left(  k_{x},k_{y}\right)  $. $k_{x}=k_{F}\cos\phi$ and
$k_{y}=k_{F}\sin\phi$ are the wave vector components outside the scattering
region and kinetic energy $E-V_{\sigma}=\hbar^{2}(q_{x}^{\sigma2}%
+q_{y}^{\sigma2})/\left(  2m^{\ast}\right)  $. The boundary conditions at the
potential steps are $\psi_{I,\sigma}(0,y)=\psi_{II,\sigma}(0,y)$,
$\psi_{II,\sigma}(D,y)=\psi_{III,\sigma}(D,y)$ and $k_{y}=q_{y}^{\sigma}$.
Except for the singular point $V_{\sigma}=E-\hbar^{2}k_{y}^{2}/(2m^{\ast})$
(or $q_{x}^{\sigma2}=0$), the transmission and reflection coefficients read
\begin{gather}
t_{\mathbf{k}}^{\sigma}=\frac{2k_{x}q_{x,\mathbf{{k}}}^{\sigma}e^{-ik_{x}D}%
}{i(k_{x}^{2}+q_{x}^{\sigma2})\sin q_{x,\mathbf{{k}}}^{\sigma}D+2k_{x}%
q_{x,\mathbf{{k}}}^{\sigma}\cos q_{x,\mathbf{{k}}}^{\sigma}D}\\
r_{\mathbf{k}}^{\sigma}=\frac{i(q_{x}^{\sigma2}-k_{x}^{2})\sin
q_{x,\mathbf{{k}}}^{\sigma}D}{i(k_{x}^{2}+q_{x}^{\sigma2})\sin
q_{x,\mathbf{{k}}}^{\sigma}D+2k_{x}q_{x,\mathbf{{k}}}^{\sigma}\cos
q_{x,\mathbf{{k}}}^{\sigma}D}.
\end{gather}
When $E\gg V_{\sigma}$, i.e. when the potential steps are relatively small,
\textbf{GB}: \textit{Can you please check changes?}
\[
q_{x}^{\sigma}=\pm\sqrt{\frac{2m^{\ast}}{\hbar^{2}}\left(  E-V_{\sigma
}\right)  -q_{y}^{\sigma2}}\rightarrow\pm\sqrt{\frac{2m^{\ast}}{\hbar^{2}%
}E-k_{y}^{2}}=\pm k_{x}.
\]
Also $q_{x,\mathbf{{k}}}^{\sigma}\approx\sqrt{-2m^{\ast}V_{\sigma}}/\hbar
\gg\sqrt{2m^{\ast}E}/\hbar=k$ and we obtain the simplified expressions
\begin{align}
t_{\mathbf{k}}^{\sigma}  &  \cong\frac{2k_{F}e^{-ikD\cos\phi}\cos\phi
}{iq_{x,\mathbf{{k}}}^{\sigma}\sin q_{x,\mathbf{{k}}}^{\sigma}D+2k_{F}\cos
\phi\cos q_{x,\mathbf{{k}}}^{\sigma}D}\nonumber\\
&  \cong\left\{
\begin{array}
[c]{ll}%
e^{-ik_{F}D\cos\phi}, & \quad q_{x,\mathbf{{k}}}^{\sigma}D=2n\pi\\
-e^{-ik_{F}D\cos\phi}, & \quad q_{x,\mathbf{{k}}}^{\sigma}D=(2n+1)\pi\\
0, & \quad q_{x,\mathbf{{k}}}^{\sigma}D\neq n\pi
\end{array}
\right. \\
r_{\mathbf{k}}^{\sigma}  &  \cong\frac{iq_{x,\mathbf{{k}}}^{\sigma}\sin
q_{x,\mathbf{{k}}}^{\sigma}D}{iq_{x,\mathbf{{k}}}^{\sigma}\sin
q_{x,\mathbf{{k}}}^{\sigma}D+2k_{F}\cos\phi\cos q_{x,\mathbf{{k}}}^{\sigma}%
D}\nonumber\\
&  \cong\left\{
\begin{array}
[c]{ll}%
0, & \quad q_{x,\mathbf{{k}}}^{\sigma}D=n\pi\\
1, & \quad q_{x,\mathbf{{k}}}^{\sigma}D\neq n\pi
\end{array}
.\right.
\end{align}
For the special point $V^{\sigma}=E-\hbar^{2}k_{y}^{2}/\left(  2m^{\ast
}\right)  $, the transmission and reflection coefficients reduce to \textit{
}
\begin{equation}
t_{\mathbf{k}}^{\sigma}=\frac{-2ie^{-ik_{x}D}}{ik_{x}D-2},r_{\mathbf{k}%
}^{\sigma}=\frac{ik_{x}D}{ik_{x}D-2}.
\end{equation}

Next we consider the limit of a slowly varying potential $V_{\sigma
}^{(\mathrm{slow})}$as been defined before Eq. (\ref{Vapr}):
\begin{equation}
H^{\mathrm{2DEG}}=-\frac{\hbar^{2}\nabla^{2}}{2m^{\ast}}+V_{\sigma
}^{(\mathrm{slow})}(\mathbf{r}).
\end{equation}
If $V_{\sigma}$ is smaller than $E_{x,\mathbf{k}}(\equiv E_{F}-\hbar^{2}%
k_{y}^{2}/2m^{\ast})$, the wave function can be written as
\begin{equation}
\psi_{\sigma,\mathbf{k}}(x,y)=\left(  c^{\sigma}e^{i\int_{x_{0}}%
^{x}q_{x,\mathbf{{k}}}^{\sigma}(x^{\prime})dx^{\prime}}+d^{\sigma}%
e^{-i\int_{x_{0}}^{x}q_{x,\mathbf{{k}}}^{\sigma}(x^{\prime})dx^{\prime}%
}\right)  e^{ik_{y}y}.
\end{equation}
where $q_{x,\mathbf{{k}}}^{\sigma}(x^{\prime})\equiv\sqrt{2m^{\ast}%
(E_{F}-V_{\sigma}(x^{\prime2}k_{y}^{2}/2m^{\ast})}/\hbar$ and $x_{0}$ is a
reference point. In region III
\begin{align}
\psi_{III,\sigma,\mathbf{k}}(x,y)  &  =t^{\sigma}e^{ik_{x}x+ik_{y}%
y}\nonumber\\
&  =\left(  e^{i\int_{0}^{x}q_{x,\mathbf{{k}}}^{\sigma}(x^{\prime})dx^{\prime
}}+r^{\sigma}e^{-i\int_{0}^{x}q_{x,\mathbf{{k}}}^{\sigma}(x^{\prime
})dx^{\prime}}\right)  e^{ik_{y}y}\nonumber\\
&  \cong\left(  e^{ik_{x}x+i(q_{x,\mathbf{{k}}}^{\sigma}-k_{x})D}+r^{\sigma
}e^{-ik_{x}x-i(q_{x,\mathbf{{k}}}^{\sigma}-k_{x})D}\right)  e^{ik_{y}y}.
\end{align}
In the semiclassical approximation electrons cannot pass the scattering region
when the potential energy is larger than its kinetic energy $E_{x,\mathbf{k}}%
$. This leads to the transmission and reflection coefficients
\begin{align}
t_{\mathbf{k}}^{\sigma}  &  =\left\{
\begin{array}
[c]{ll}%
0, & \quad E_{x,\mathbf{k}}<V_{\sigma}\\
e^{i(q_{x,\mathbf{{k}}}^{\sigma}-k_{x})D}, & \quad E_{x,\mathbf{k}}>V_{\sigma}%
\end{array}
\right. \\
r_{\mathbf{k}}^{\sigma}  &  =\left\{
\begin{array}
[c]{ll}%
1, & \quad E_{x,\mathbf{k}}<V_{\sigma}\\
0, & \quad E_{x,\mathbf{k}}>V_{\sigma}%
\end{array}
.\right.
\end{align}
where we kept the phase of the transmitted electron waves. Note that the
phases accumulated by the adiabatic rise and drop of the potential outside the
gate cancel eachother. \begin{figure}[th]
\begin{center}
\includegraphics[width=0.475\textwidth]{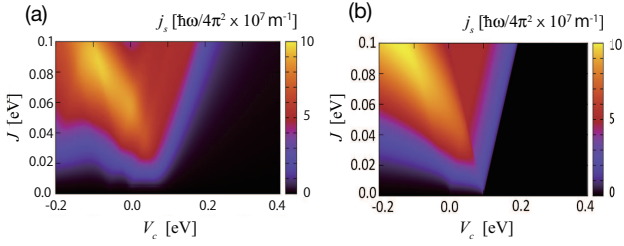}
\end{center}
\caption{Spin current per unit width pumped into a 2DEG by an electrically
insulating ferromagnetic top contact as a function of the gate voltage $V_{c}$
and the proximity exchange coupling constant $J,$ assuming that (a) the
potential at the contact edges is abrupt and (b) continuous. The chemical
potential of the 2DEG $E_{F}=0.1\,\,$eV. The transverse length of the top
contact is $D=10\,\,\,$nm. }%
\label{schpump}%
\end{figure}

We display the computed spin current $j_{s}$ pumped into a 2DEG for an abrupt
potential in Fig. \ref{schpump} (a) and for a slowly varying potential in Fig.
\ref{schpump} (b) as a function of the Fermi energy $E_{F}$, the gate voltage
$V_{c}$, exchange coupling $J,$ and length $D$. A larger exchange splitting
$J$ increases the spin current pumped into the 2DEG, as expected in the limit
of a weak ferromagnet. When $V_{\sigma}>E_{F}$, the wave function with spin
$\sigma$ in the limit of an abrupt potential exponentially decays under the
FI. The white-dashed line in Fig. \ref{schpump}(a) is the function
$V_{\uparrow}=V_{c}-J=E_{F}.$ To the far right of it, $V_{c}\gg E_{F}$ and
$j_{s}$ vanishes since all electrons are reflected by the high potential
barrier. For the slowly varying potentials, electrons are completely reflected
when $V_{\uparrow,\downarrow}>E_{F}$ and
$j_{s}$ vanishes abruptly at the same line as seen in Fig. \ref{schpump}(b).
When $V_{\sigma}<E_{F}$, electron waves may interfere constructively to
maximize the spin current, forming the broad ridge of enhanced spin currents
that is observed for both potentials. The black dashed line in Fig.
\ref{schpump}(a) is a guide to the eye that coincides with the maximum spin
pumping current which occurs when the phase difference between of up spin and
down spin electrons becomes large.

In general, we observe that at least for the considered parameter regime, a
WKB-like approximation of the spin pumping into the 2DEG that is valid for
slowly varying potentials agrees well with the fully quantum mechanical result
for abrupt potentials.

\subsection{Graphene}

We now turn to spin pumping into graphene, focusing on the $\mathbf{K}$ valley
and multiply the result by 2, thereby disregarding intervalley scattering. As
before, the electrons in graphene experience the proximity-exchange and
electrostatic potentials collected in $U(\mathbf{r},t)$. We consider the
parameter regimes $|E_{F}|,|V_{\sigma}|<1\,$eV, for which the standard
envelope wave function Hamiltonian with energy zero at the Dirac point
applies:%
\begin{equation}
H_{kin}^{\mathrm{graphene}}=-i\hbar v_{F}\boldsymbol{\hat{\sigma}}%
\cdot\mathbf{\nabla},
\end{equation}
where $\boldsymbol{\hat{\sigma}}=\sum_{l}\hat{\sigma}_{l}\mathbf{e}_{l}$ and
$\hat{\sigma}_{l}$ ($l=x,y$) are the Pauli matrices in pseudo-spin space. The
envelope wave function is the spinor
\begin{equation}
\psi\left(  \mathbf{r}\right)  =\left(
\begin{array}
[c]{c}%
\psi_{1}\left(  \mathbf{r}\right) \\
\psi_{2}\left(  \mathbf{r}\right)
\end{array}
\right)  .
\end{equation}

We discuss again abrupt potentials first:
\begin{equation}
H^{\mathrm{graphene}}=-i\hbar v_{F}\boldsymbol{\hat{\sigma}}\cdot
\mathbf{\nabla}+V_{\sigma}^{\left(  \mathrm{abrupt}\right)  }(\mathbf{{r}).}%
\end{equation}
In region I ($x<0$)
\begin{align}
\psi_{I,\sigma,\mathbf{k}}\left(  \mathbf{r}\right)   &  =\left(
\begin{array}
[c]{c}%
1\\
\chi e^{i\theta_{\mathbf{k}}}%
\end{array}
\right)  e^{ik_{x}x+ik_{y}y}\nonumber\\
&  +r_{\mathbf{k}}^{\sigma}\left(
\begin{array}
[c]{c}%
1\\
-\chi e^{-i\theta_{\mathbf{k}}}%
\end{array}
\right)  e^{-ik_{x}x+ik_{y}y},
\end{align}
while in region II ($0<x<D$) \textit{ }%
\begin{align}
\psi_{II,\sigma,\mathbf{k}}\left(  \mathbf{r}\right)   &  =a_{\mathbf{k}%
}^{\sigma}\left(
\begin{array}
[c]{c}%
1\\
\chi^{\sigma}e^{i\theta_{\mathbf{q}_{\sigma}}}%
\end{array}
\right)  e^{iq_{x,\mathbf{{k}}}^{\sigma}x+iq_{y,\sigma}y}\nonumber\\
&  +b_{\mathbf{k}}^{\sigma}\left(
\begin{array}
[c]{c}%
1\\
-\chi^{\sigma}e^{-i\theta_{\mathbf{q}_{\sigma}}}%
\end{array}
\right)  e^{-iq_{x,\mathbf{{k}}}^{\sigma}x+iq_{y,\sigma}y},
\end{align}
where $q_{x,\mathbf{{k}}}^{\sigma}=\sqrt{(E-V_{\sigma})^{2}/\hbar^{2}v_{F}%
^{2}-k_{y}^{2}}.$In region III ($x>D$)
\begin{equation}
\psi_{III,\sigma,\mathbf{k}}\left(  \mathbf{r}\right)  =t_{\mathbf{k}}%
^{\sigma}\left(
\begin{array}
[c]{c}%
1\\
\chi e^{i\theta_{\mathbf{k}}}%
\end{array}
\right)  e^{ik_{x}x+ik_{y}y},
\end{equation}
where $\chi=\mathrm{sgn}(E)$, $\chi^{\sigma}=\mathrm{sgn}\left(  E-V_{\sigma
}\right)  $, $\tan\theta_{\mathbf{k}}=k_{y}/k_{x}$, $\tan\theta_{\mathbf{q}%
_{\sigma}}=q_{y,\sigma}/q_{x,\mathbf{{k}}}^{\sigma}$. The boundary conditions
are $\psi_{I,\sigma}(0,y)=\psi_{II,\sigma}(0,y),\,\psi_{II,\sigma}%
(D,y)=\psi_{III,\sigma}(D,y)$ and $k_{y}=q_{y,\sigma}.$ When $\left\vert
E-V^{\sigma}\right\vert >\hbar v_{F}|k_{y}|$ (propagating states in the gated
region) transmission and reflection coefficients read%
\begin{align}
t_{\mathbf{k}}^{\sigma}(\phi)  &  =\frac{e^{-ik_{x}D}\cos\theta_{\sigma}%
\cos\phi}{X_{\mathbf{k}}^{\sigma}}\\
r_{\mathbf{k}}^{\sigma}(\phi)  &  =\frac{ie^{i\phi}(\chi\chi^{\sigma}\sin
\phi-\sin\theta_{\sigma})\sin q_{x,\mathbf{{k}}}^{\sigma}D}{X_{\mathbf{k}%
}^{\sigma}},
\end{align}
where $X_{\mathbf{k}}^{\sigma}\equiv\cos\theta_{\sigma}\cos\phi\cos
q_{x,\mathbf{{k}}}^{\sigma}D+i(\sin\theta_{\sigma}\sin\phi-\chi\chi^{\sigma
})\sin q_{x,\mathbf{{k}}}^{\sigma}D$.

When $\left\vert E-V^{\sigma}\right\vert <\hbar v_{F}|k_{y}|$ (evanescent
states in the gated region), we substitute $q_{x,\mathbf{{k}}}^{\sigma
}=i\kappa_{x,\sigma}$, $e^{i\theta_{\sigma}}\rightarrow(i\kappa_{x,\sigma
}+ik_{y})/(\left\vert E-V\right\vert /\hbar v_{F})$ and $-e^{-i\theta_{\sigma
}}\rightarrow(-i\kappa_{x,\sigma}+ik_{y})(|E-V|/\hbar v_{F})$ for propagating
states. Transmission and reflection coefficients then become
\begin{gather}
t_{\mathbf{k}}^{\sigma}(\phi)=\frac{i\chi\hbar v_{F}\kappa_{x,\sigma}\cos\phi
e^{-ik_{F}D\cos\phi}}{Y_{\mathbf{k}}^{\sigma}}\\
r_{\mathbf{k}}^{\sigma}(\phi)=\frac{-iV_{\sigma}e^{i\phi}\sinh\kappa
_{x,\sigma}D\sin\phi}{Y_{\mathbf{k}}^{\sigma}}.
\end{gather}
where $Y_{\mathbf{k}}^{\sigma}\equiv(E\cos^{2}\phi-V_{\sigma})\sinh
\kappa_{x,\sigma}D+i\chi\hbar v_{F}\kappa_{x,\sigma}\cos\phi\cosh
\kappa_{x,\sigma}D$.

When $E=V_{\sigma}$, the wave function in region II can be written as
\begin{equation}
\psi_{II,\sigma,\mathbf{k}}\left(  \mathbf{r}\right)  =\left(
\begin{array}
[c]{c}%
Ae^{q_{y}(x+iy)}\\
Be^{\kappa_{y}(-x+iy)}%
\end{array}
\right)  .
\end{equation}
With specular scattering boundary condition $q_{y}=\kappa_{y}=k_{y}$:
\begin{gather}
t_{\mathbf{k}}^{\sigma}(\phi)=\frac{2e^{-ik_{y}D}\cos\phi}{e^{i\phi}e^{k_{y}%
D}+e^{-i\phi}e^{-k_{y}D}},\\
r_{\mathbf{k}}^{\sigma}(\phi)=\frac{-2e^{i\phi}\sinh k_{y}D}{e^{i\phi}%
e^{k_{y}D}+e^{-i\phi}e^{-k_{y}D}}.
\end{gather}
When $\left\vert E\right\vert \ll\left\vert V_{\sigma}\right\vert ,$ i.e. the
Fermi circle under the gate is much larger than that in the leads, $\cos
\theta_{\sigma}\rightarrow1$, $\sin\theta_{\sigma}\rightarrow0$, and
$q_{x,\mathbf{{k}}}^{\sigma}\cong\left\vert V_{\sigma}\right\vert /\hbar
v_{F}=\left\vert V_{c}\mp J\right\vert /\hbar v_{F}$, leading to the
simplified results
\begin{gather}
t_{\mathbf{k}}^{\sigma}(\phi)\rightarrow\frac{\cos\phi e^{-ik_{F}D\cos\phi}%
}{\cos\phi\cos q_{x,\mathbf{{k}}}^{\sigma}D-i\chi\chi^{\sigma}\sin
q_{x,\mathbf{{k}}}^{\sigma}D},\\
r_{\mathbf{k}}^{\sigma}(\phi)\rightarrow\frac{i\chi\chi^{\sigma}e^{i\phi}%
\sin\phi\sin q_{x,\mathbf{{k}}}^{\sigma}D}{\cos\phi\cos q_{x,\mathbf{{k}}%
}^{\sigma}D-i\chi\chi^{\sigma}\sin q_{x,\mathbf{{k}}}^{\sigma}D}.
\end{gather}

We now turn to a slowly varying potential with Hamiltonian
\begin{equation}
H^{\mathrm{graphene}}=-i\hbar v_{F}\boldsymbol{\hat{\sigma}}\cdot
\mathbf{\nabla}+V_{\sigma}^{\left(  \mathrm{slow}\right)  }\left(
\mathbf{r}\right)  .
\end{equation}
If $V_{\sigma}$ is smaller than $E_{x,\mathbf{k}}(\equiv E_{F}-\hbar
v_{F}|k_{y}|)$, the wave function can be written as
\begin{align}
\psi_{\sigma,\mathbf{k}}\left(  \mathbf{r}\right)   &  =c^{\sigma}\left(
\begin{array}
[c]{c}%
G_{\sigma,\mathbf{k}}^{-1/2}+iG_{\sigma,\mathbf{k}}^{1/2}k_{y}/|k_{y}|\\
G_{\sigma,\mathbf{k}}^{-1/2}-iG_{\sigma,\mathbf{k}}^{1/2}k_{y}/|k_{y}|
\end{array}
\right)  e^{i\int_{x_{0}}^{x}q_{x,\mathbf{{k}}}^{\sigma}(x^{\prime}%
)dx^{\prime}+ik_{y}y}\nonumber\\
&  +d^{\sigma}\left(
\begin{array}
[c]{c}%
G_{\sigma,\mathbf{k}}^{1/2}+iG_{\sigma,\mathbf{k}}^{-1/2}k_{y}/|k_{y}|\\
G_{\sigma,\mathbf{k}}^{1/2}-iG_{\sigma,\mathbf{k}}^{-1/2}k_{y}/|k_{y}|
\end{array}
\right)  e^{-i\int_{x_{0}}^{x}q_{x,\mathbf{{k}}}^{\sigma}(x^{\prime
})dx^{\prime}+ik_{y}y},
\end{align}
where $x_{0}$ is a reference point, $q_{x,\mathbf{{k}}}^{\sigma}=\left\vert
E_{x,\mathbf{k}}-V_{\sigma}(x)\right\vert /\hbar v_{F}$, and with $\nu
\equiv\mathrm{sgn}\left(  V_{\sigma}(x_{0})-E\right)  $
\begin{equation}
G_{\sigma,\mathbf{k}}=\left(  \frac{|E-V_{\sigma}(x)|/\hbar v_{F}+k_{x}%
(x)}{|k_{y}|}\right)  ^{\nu}.
\end{equation}
In region III:
\begin{align}
\psi_{III,\sigma,\mathbf{k}}  &  \left(  \mathbf{r}\right)  =t^{\sigma}\left(
\begin{array}
[c]{c}%
G_{\sigma,\mathbf{k}}^{-1/2}+iG_{\sigma,\mathbf{k}}^{1/2}k_{y}/|k_{y}|\\
G_{\sigma,\mathbf{k}}^{-1/2}-iG_{\sigma,\mathbf{k}}^{1/2}k_{y}/|k_{y}|
\end{array}
\right)  e^{ik_{x}x+ik_{y}y}\nonumber\\
&  =\left(
\begin{array}
[c]{c}%
G_{\sigma,\mathbf{k}}^{-1/2}+iG_{\sigma,\mathbf{k}}^{1/2}k_{y}/|k_{y}|\\
G_{\sigma,\mathbf{k}}^{-1/2}-iG_{\sigma,\mathbf{k}}^{1/2}k_{y}/|k_{y}|
\end{array}
\right)  e^{i\int_{x_{0}}^{x}q_{x,\mathbf{{k}}}^{\sigma}(x^{\prime}%
)dx^{\prime}+ik_{y}y}\nonumber\\
&  +r^{\sigma}\left(
\begin{array}
[c]{c}%
G_{\sigma,\mathbf{k}}^{1/2}+iG_{\sigma,\mathbf{k}}^{-1/2}k_{y}/|k_{y}|\\
G_{\sigma,\mathbf{k}}^{1/2}-iG_{\sigma,\mathbf{k}}^{-1/2}k_{y}/|k_{y}|
\end{array}
\right)  e^{-i\int_{x_{0}}^{x}q_{x,\mathbf{{k}}}^{\sigma}(x^{\prime
})dx^{\prime}+ik_{y}y}\nonumber\\
&  \cong\left(
\begin{array}
[c]{c}%
G_{\sigma,\mathbf{k}}^{-1/2}+iG_{\sigma,\mathbf{k}}^{1/2}k_{y}/|k_{y}|\\
G_{\sigma,\mathbf{k}}^{-1/2}-iG_{\sigma,\mathbf{k}}^{1/2}k_{y}/|k_{y}|
\end{array}
\right)  e^{ik_{x}x+i(q_{x,\mathbf{{k}}}^{\sigma}-k_{x})D+ik_{y}y}\nonumber\\
&  +r^{\sigma}\left(
\begin{array}
[c]{c}%
G_{\sigma,\mathbf{k}}^{1/2}+iG_{\sigma,\mathbf{k}}^{-1/2}k_{y}/|k_{y}|\\
G_{\sigma,\mathbf{k}}^{1/2}-iG_{\sigma,\mathbf{k}}^{-1/2}k_{y}/|k_{y}|
\end{array}
\right)  e^{-ik_{x}x-i(q_{x,\mathbf{{k}}}^{\sigma}-k_{x})D+ik_{y}y}.
\end{align}
In the semiclassical WKB approximation electrons cannot transmit the
scattering region when the potential energy is larger than $E_{x,\mathbf{k}}$.
This corresponds to disregarding the evanescent wave-tunneling through the
gate region. The transport coefficients then read
\begin{align}
t_{\mathbf{k}}^{\sigma}=  &  \left\{
\begin{array}
[c]{ll}%
0, & \quad E_{x,\mathbf{k}}<V_{\sigma}\\
e^{i(q_{x,\mathbf{{k}}}^{\sigma}-k_{x})D}, & \quad E_{x,\mathbf{k}}>V_{\sigma}%
\end{array}
\right. \\
r_{\mathbf{k}}^{\sigma}=  &  \left\{
\begin{array}
[c]{ll}%
1, & \quad E_{x,\mathbf{k}}<V_{\sigma}\\
0, & \quad E_{x,\mathbf{k}}>V_{\sigma}%
\end{array}
.\right.
\end{align}
where we again preserved the phase of the transmitted electron waves.
\begin{figure}[th]
\begin{center}
\includegraphics[width=0.475\textwidth]{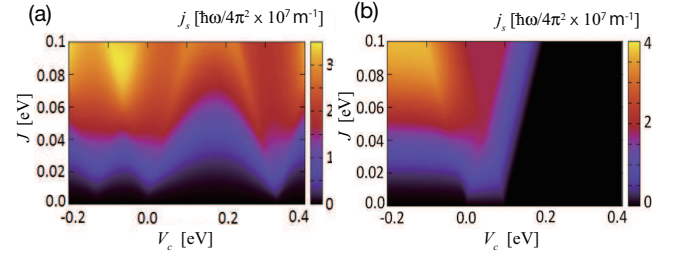}
\end{center}
\caption{(a) Spin current per unit width pumped into graphene by a
ferromagnetic top layer as a function of the gate voltage $V_{c}$ and the
proximity exchange coupling constant $J$, assuming that (a) the potential at
the contact/gate edges is abrupt and (b) slowly varying. The (zero-bias)
chemical potential of the graphene $\mu=100\,$ meV. The length of the
ferromagnetic region is $D=10\,$nm. $m=0.9$ is the cosine of the magnetization
precession cone angle. }%
\label{grapump}%
\end{figure}

\begin{figure}[th]
\begin{center}
\includegraphics[width=0.475\textwidth]{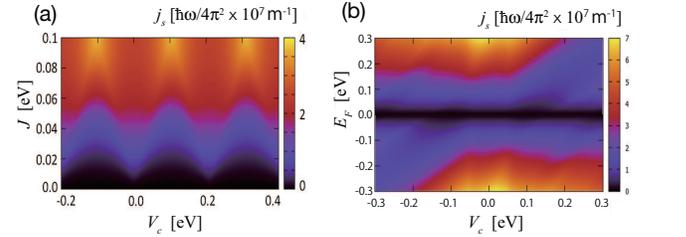}
\end{center}
\caption{ (a) Spin current per unit width as in Fig. \ref{grapump} but for
$\mu=1\,\,$meV, i.e. close to the Dirac electron neutrality point. (b) Spin
current density pumped into graphene as a function of the (non-adiabatic) gate
voltage $V_{c}$ and Fermi energy of the reservoirs $E_{F}$. Other parameters
are $J=50\,\,\,$meV, gate length $D=10\,\,$nm, and cosine of the magnetization
precession cone angle $m=0.9$. }%
\label{grasym}%
\end{figure}

We plot the dependence of the spin current density $j_{s}$ pumped into
graphene on $V_{c}$ and $J$ for an abrupt potential at the contact/gate edges
in Fig. \ref{grapump} (a) and a slowly varying potential in Fig. \ref{grapump}
(b).
These figures can be compared with the 2DEG device for the same parameters.
In the 2DEG $j_{s}$ vanishes when $V_{\sigma}>E_{F}$ because all electrons are
reflected by the potential barrier. On the other hand, in graphene
$j_{s}$ does not vanish even when $V_{\sigma}>E_{F}$, because electrons can
propagate through the potential by Klein tunneling via the valence band
states. In the abrupt potential limit, when $\left(  E_{F}-V_{\sigma}\right)
/E_{F}<-1$ electrons are seen to tunnel efficiently through the gate region.
This renders the physics of transport including spin pumping in graphene very
different from that of the 2DEG. For the adiabatic potential, however,
electrons waves are reflected and Klein tunneling \cite{reijnders2013} does
not occur. Therefore, when $V_{\uparrow}>E_{F}$,
$j_{s}$ vanishes.

In Fig. \ref{grasym}(a), we plot $j_{s}$ for graphene close to
electroneutrality ($E_{F}=1\,\mathrm{meV}$) , while in Fig. \ref{grapump} (a)
the Fermi energy is substantial. Note that we do not address the complications
of transport at the Dirac point (see e.g. \cite{Chen2008}). Since the electron
density is very small, the spin pumping spectra in Fig. \ref{grasym}(b)
reflect the particle-hole symmetry of the graphene band structure. The spin
current in n%
$\vert$%
n(p)-F%
$\vert$%
n-graphene is the same as that in p%
$\vert$%
p(n)-F%
$\vert$%
p graphene junctions summarized by the general symmetry relation%
\begin{equation}
j_{s}(E_{F},V_{c})=j_{s}(-E_{F},-V_{c})
\end{equation}
We observe in Fig.\ref{grasym}(b) that in the area between the liness
$V_{c}\pm E_{F}$ the spin current is suppressed because the modes under the
gate are evanescent.

\section{Conclusion}

In conclusion, we report spin pumping into a 2DEG and graphene by a planar
contact consisting of a magnetic insulator film with a metal gate. In both
cases the spin current can be controlled by the gate voltage that modulates
the electron density. The pumped spin currents in both systems are remarkably
different in the abrupt potential limit, reflecting the difference between the
Schr\"{o}dinger and Dirac equations. Rahimi and Moghaddam \cite{rahimi2014}
compute spin pumping into graphene for the same configuration. They do not
discuss the 2DEG and focus on different parameter regimes, however.\textit{
}The present theory is valid when the scattering mean-free path is smaller
than the system size. We therefore chose a narrow gate with $D=10\,$nm. In
contrast to perturbation theory are no restrictions on the magnitude and cone
angle of the induced exchange potential. Even for a large $J=50\,\mathrm{meV}%
$, the coherence length is $\lambda\simeq20\,$nm, which means that we are in
the limit of a weak ferromagnet. We predict typical dependence of the spin
current on all device parameters that can be very different for graphene and
the the 2DEG, mainly by Klein tunneling in the former. The effects can in
principle be observed by metallic contacts outside the gated region or
enhanced broadening of the ferromagnetic resonance, which requires additional
but straightforward modelling of the specific sample.

Future work should take into account disorder scattering, including spin flip
scattering, that has been found to be negligible when graphene has a large
contact area with a ferromagnet.

\acknowledgements This work was supported by JSPS KAKENHI Grant Nos. 25247056,
25220910, 26103006, JP26400308, and JP15H05854. G.B acknowledges the hospitality of the Zernike Institute
of the University of Groningen and useful discussions with Bart van Wees and
Christian Leutenantsmeyer.


\end{document}